\documentclass[conference]{IEEEtran}
\IEEEoverridecommandlockouts

\usepackage{hyperref}
\usepackage{cite}
\usepackage{amsmath,amssymb,amsfonts}
\usepackage{algorithmic}
\usepackage{graphicx}
\usepackage{textcomp}
\usepackage{xcolor}
\def\BibTeX{{\rm B\kern-.05em{\sc i\kern-.025em b}\kern-.08em
    T\kern-.1667em\lower.7ex\hbox{E}\kern-.125emX}}

\definecolor{arytenoid-cartilage}{rgb}{0.54, 0.17, 0.89} 
\definecolor{epiglottis}{rgb}{0.25, 0.88, 0.82} 
\definecolor{lower-lip}{rgb}{0.00, 1.00, 0.00} 
\definecolor{pharyngeal-wall}{rgb}{0.85, 0.66, 0.13} 
\definecolor{soft-palate-midline}{rgb}{0.12, 0.56, 1.00} 
\definecolor{tongue}{rgb}{1.00, 0.55, 0.00} 
\definecolor{upper-lip}{rgb}{1.00, 0.00, 1.00} 
\definecolor{vocal-folds}{rgb}{1.00, 0.41, 0.71} 
\begin{document}

\title{Reconstruction of the Vocal Tract from Speech via Phonetic Representations Using MRI Data\\}

\author{\IEEEauthorblockN{Sofiane Azzouz}
\IEEEauthorblockA{\textit{Multispeech} \\
\textit{Université de Lorraine, CNRS, Inria, F-54000}\\
Nancy, France \\
sofiane.azzouz@loria.fr}
\and
\IEEEauthorblockN{Pierre-André Vuissoz}
\IEEEauthorblockA{\textit{IADI} \\
\textit{Université de Lorraine, Inserm, IADI U1254, F-54000}\\
Nancy, France \\
pa.vuissoz@chru-nancy.fr}
\and
\IEEEauthorblockN{Yves Laprie}
\IEEEauthorblockA{\textit{Multispeech} \\
\textit{Université de Lorraine, CNRS, Inria, F-54000}\\
Nancy, France \\
yves.laprie@loria.fr}
}


\maketitle

\begin{abstract}
Articulatory acoustic inversion aims to reconstruct the complete geometry of the vocal tract from the speech signal. In this paper, we present a comparative study of several levels of phonetic segmentation accuracy, together with a comparison to the baseline introduced in our previous work, which is based on Mel-Frequency Cepstral Coefficients (MFCCs). All the approaches considered are based on a denoised speech signal and aim to investigate the impact of incorporating phonetic information through three successive levels: an uncorrected automatic transcription, a temporally aligned phonetic segmentation, and an expert manual correction following alignment. The models are trained to predict articulatory contours extracted from vocal tract MRI images using an automatic contour tracking method. The results show that, among the models relying on phonetic representations, manual correction after alignment yields the best performance, approaching that of the baseline.
\end{abstract}


\section{Introduction}

Acoustic-to-articulatory inversion aims to reconstruct the shape of the vocal tract from the audio signal. Several approaches have been proposed to address this inverse problem, relying on different types of data and methodologies.
Initially, available data were limited and consisted mainly of static X-ray images and area functions. Early methods were based on physical approaches, using acoustic simulations and simplified approximations of the vocal tract. For instance, the Coker model\cite{coker2005model} represents an early attempt at articulatory modeling applied to speech synthesis.

Subsequently, pioneering works such as those by \cite{atal1970determination, wakita1973} addressed inversion using acoustic transfer functions. In particular, \cite{atal1978inversion} demonstrated that the problem is fundamentally ambiguous, with an infinite number of possible solutions for a given formant triplet.

The emergence of richer data, such as cineradiography (X-ray films), enabled the exploration of new approaches based on articulatory models. The model proposed by \cite{maeda1982digital}, built from 500 images, represented a major advancement in speech synthesis and articulatory phonetics research. This model was later used for inversion by formulating the problem as an ill-posed inverse problem, often solved through regularization and optimization techniques.

In the 1990s, \cite{rahim1990estimation} introduced the use of neural networks to approximate solutions to the acoustic-to-articulatory inversion problem. In parallel, codebook-based approaches were explored \cite{schroeter1994techniques, Ouni2005}, which consist of searching an acoustic–articulatory dictionary for entries closest to the signal to be inverted and then interpolating the corresponding articulatory configurations.

With the availability of articulatory datasets acquired using Electromagnetic Articulography (EMA) and X-ray microbeam systems, such as the mngu0 corpus \cite{richmond2011announcing}, it became possible to establish a direct relationship between acoustic signals and articulatory geometry without relying on explicit physical models \cite{katsamanis2008inversion}. These datasets fostered the development of data-driven statistical methods, including approaches based on Gaussian Mixture Models (GMMs) and Hidden Markov Models (HMMs), as well as neural network–based techniques. In particular, recurrent neural architectures have been widely explored, including Bi-LSTM models \cite{liu2015deep, parrot2020}. In addition, \cite{biasutto2018phoneme} proposed a Bi-GRU–based approach for phoneme-to-articulatory mapping using EMA data. More complex architectures, such as CNN–BiLSTM \cite{shahrebabaki2020sequence} and Bi-GRNN \cite{wu2023speaker}, as well as more recent Temporal Convolutional Networks (TCNs) \cite{siriwardena2023secret}, have also been investigated.

To further improve performance, phonetic segmentation has also been incorporated as an additional input \cite{wang2023two}. However, this strategy requires phonetic segmentation to be available at inference time, which limits its applicability. To address this issue while still benefiting from phonetic information, multitask learning frameworks have been proposed. For instance, \cite{siriwardena22_interspeech} jointly predict phonemic labels and articulatory variables, allowing the model to exploit phonetic structure without explicit segmentation during inference.

However, EMA data present several limitations: a limited number of sensors (three or four on the tongue), the rigidity introduced by the glue used to attach these sensors (particularly on the tongue and lips), and the presence of wires between the lips, which can slightly disturb articulation.

More recently, real-time dynamic MRI (rt-MRI) has been introduced as an alternative, with the first recorded corpus described in \cite{ramanarayanan2018analysis}. Deep learning approaches have been applied to these data \cite{Csapo2020}. Nevertheless, the use of rt-MRI remains limited due to several constraints: difficulty in acquiring sufficiently large datasets, low signal quality after denoising, the lack of robust contour-tracking tools, and relatively low spatial resolution (approximately 68×68 pixels), in addition to MRI-related artifacts.
\cite{ribeiro2022automatic} introduces the first machine learning–based approach capable of generating complete vocal tract shapes, covering all articulators and static structures from the glottis to the lips, for arbitrary phoneme sequences. By combining highly accurate phonetic segmentation, reliable contour tracking from real-time MRI data.

In this study, we used a higher-resolution rt-MRI dataset (136×136 pixels), already denoised and of better quality than most existing rt-MRI databases. Furthermore, instead of directly inverting full rt-MRI images, we opted for automatic contour tracking to extract the shapes of the different articulators, and we performed inversion using only these contours rather than the entire image.
The objective of this work is to compare phonetic segmentations at different levels of transcription accuracy—ranging from manually corrected alignments, to automatically aligned, to raw transcriptions—against the direct use of the speech signal represented by MFCCs. We aim to evaluate whether investing considerable time in manual correction is justified, or if less precise alignments or transcriptions can achieve comparable performance. This comparison will help determine the trade-offs between preprocessing effort and predictive accuracy for articulatory contour modeling.

\section{Dataset}
The corpus used in this study was recorded at the Nancy University Hospital (CHRU de Nancy). It consists of speech productions from a single native French female speaker recorded over five sessions. The corpus contains approximately 3.5 hours of speech, corresponding to about 2,100 sentences. In total, it includes 153 sequences, each lasting around 80 seconds, and associated with 4,000 MRI frames.

Real-time MRI data were acquired using a Siemens Prisma 3T scanner. Each image corresponds to an 8 mm thick mid-sagittal slice and has a spatial resolution of 136 × 136 pixels, with a pixel spacing of 1.62 mm.

The audio signal was recorded using an optical microphone at a sampling rate of 16 kHz and was subsequently denoised using \cite{Ozerov2012}.
\subsection{MRI Image Preprocessing}
\begin{figure}[ht] 
  \centering
  \includegraphics[width=0.475\textwidth]{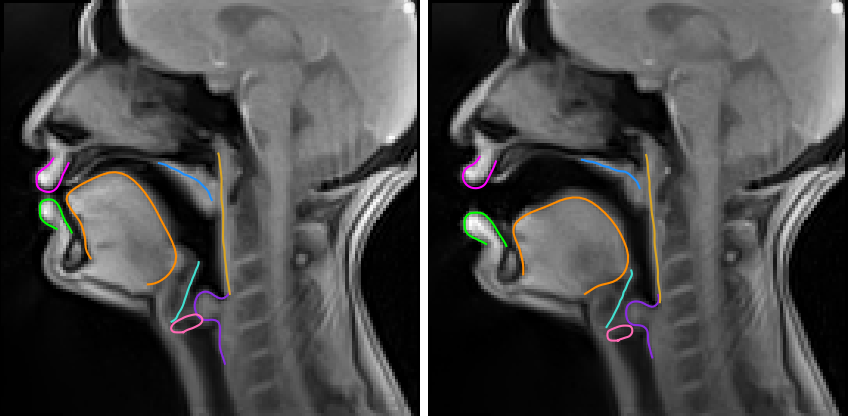} 
  \caption{
    Segmentation of articulators contour tracked in two images of the rt-MRI film:\hspace{0.2cm}
    \raisebox{0.25ex}{\textcolor{arytenoid-cartilage}{\rule{0.2cm}{0.08cm}}} Arytenoid cartilage,
    \raisebox{0.25ex}{\textcolor{epiglottis}{\rule{0.2cm}{0.08cm}}} Epiglottis,
    \raisebox{0.25ex}{\textcolor{lower-lip}{\rule{0.2cm}{0.08cm}}} Lower lip,
    \raisebox{0.25ex}{\textcolor{vocal-folds}{\rule{0.2cm}{0.08cm}}} Vocal folds,\hspace{0.12cm}
    \raisebox{0.25ex}{\textcolor{soft-palate-midline}{\rule{0.2cm}{0.08cm}}} Soft palate midline,\hspace{0.05cm}
    \raisebox{0.25ex}{\textcolor{tongue}{\rule{0.2cm}{0.08cm}}} Tongue,\hspace{0.12cm}
    \raisebox{0.25ex}{\textcolor{upper-lip}{\rule{0.2cm}{0.08cm}}} Upper lip,\hspace{0.05cm}
    \raisebox{0.25ex}{\textcolor{pharyngeal-wall}{\rule{0.2cm}{0.08cm}}} Pharyngeal wall
  }
  \label{fig:original_contours} 
\end{figure}

For MRI image preprocessing, we employed an automatic tracking approach based on a recurrent convolutional neural network (RCNN) to segment real-time MRI data into eight speech-related articulators. This segmentation procedure yields articulatory contours for each articulator, with each contour represented by 50 points. The articulators considered are the upper lip, lower lip, tongue, soft palate midline (velum), pharyngeal wall, epiglottis, arytenoid cartilage, and vocal folds , as illustrated in Figure~\ref{fig:original_contours}.

\subsection{Input preprocessing}
In our previous work \cite{azzouz2025reconstruction}, the input to the model consisted of the denoised speech signal. From this signal, we extracted the first 13 MFCC coefficients, as well as their first and second derivatives (delta and delta-delta).

In the present study, our objective is to train the model based on phonetic segmentation. We retain the denoised speech signal as the starting point. To this end, we employed three different methods.

The first method uses a pre-trained and fine-tuned Wav2Vec 2.0 model for French phonemic transcription \cite{lmssc-wav2vec2-base-phonemizer-french_2023}. This model is based on self-supervised learning and uses an output layer trained with a Connectionist Temporal Classification (CTC) loss function. For each time frame, the model produces a 61-dimensional logits vector, corresponding to the unnormalized scores associated with each phoneme in the vocabulary. These logits are then converted into phonetic probability distributions. A softmax function is applied along the phoneme dimension, producing, for each frame, a probabilistic representation of phoneme presence. To stabilize the training of the articulatory prediction model and reduce variability, these probabilities are further normalized globally on a per-session basis by subtracting the mean and dividing by the standard deviation.

The second method relies on forced alignment using the Astali tool \cite{fohr2015importance}. This approach jointly utilizes the audio signal and a temporally aligned phonetic transcription. The transcriptions are provided at the sentence level, with start and end times, and are used to perform a forced alignment between the acoustic signal and the corresponding phonemic sequence. This alignment allows precise association of each phoneme with a time interval in the audio signal. 37 distinct phonemes were obtained from this process. The extracted phonemes are then encoded as one-hot vectors, with each vector representing the presence of a single phoneme in the phonetic vocabulary.

The third and final method builds on the phonetic segmentation obtained with Astali, which was then manually corrected by an expert. The correction primarily refines the temporal boundaries between sounds. Additionally, the closure of voiceless plosives was separated from the burst, as the articulatory position differs significantly between these two parts. After this correction, 44 distinct phonemes were obtained, which were also encoded as one-hot vectors.

\section{Methods}
\subsection{Model architecture}
\begin{figure}[ht]
    \centering
    \includegraphics[width=0.47\textwidth]{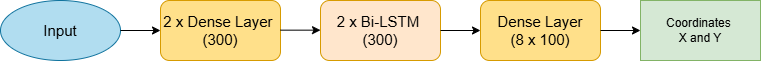} 
    \caption{Model architecture}
    \label{fig:model_architecture}
\end{figure}

The model used is the same as in our previous work \cite{azzouz2025complete, azzouz2025reconstruction}. We modified it to accept phonetic segmentation as input (see figure~\ref{fig:model_architecture}), instead of the speech signal. It consists of five layers: the first two are dense layers with 300 units each, followed by two Bi-LSTM layers with 300 units. The output layer is a dense layer with 800 (8x100) units, corresponding to the 8 articulators and their 100 respective coordinates.
\subsection{Loss function}
Since articulatory acoustic inversion is a regression task, we used mean squared error (MSE) as the loss function, which is the most commonly adopted in this context.
\begin{equation}
\text{MSE} = \frac{1}{n} \sum_{i=1}^{n} \left( y_i - \hat{y}_i \right)^2\label{eq2}
\end{equation}
where n represents the total number of observations in the dataset, \(y_i\) and \(\hat{y}_i\) represent the true and predicted values of the output for example \(i\), respectively.

\subsection{Evaluation}
We denormalized the original and predicted contours and evaluated the root mean square error (RMSE) and the median, both expressed in millimeters. For each frame, we computed the RMSE and median for each articulator, based on the 100 predicted and original points representing its geometry. Finally, we averaged the RMSE and median over all frames for each articulator, and then across all articulators.
\begin{equation}
\text{RMSE} = \sqrt{\frac{1}{n} \sum_{i=1}^{n} \left( y_i - \hat{y}_i \right)^2}\label{eq4}
\end{equation}

\subsection{Experiments}
These experiments allow us to evaluate the differences obtained between various levels of preprocessing. Using the raw speech signal directly requires no special preparation, and even the Wav2Vec 2.0 transcription does not require any human intervention. However, its precision remains limited since it is only a transcription. In contrast, using Astali  requires preparing the spoken sentences, which can be time-consuming, and manual correction demands even more effort. Our goal is to assess whether it is worthwhile to rely on phonetic segmentation or to use the denoised speech signal directly. Furthermore, we aim to determine which type of segmentation is most effective: an automatic transcription requiring no effort, forced alignment with Astali, or forced alignment followed by manual correction.

We conducted four different experiments. The first serves as a baseline, in which the model is trained using MFCCs as input; this will be referred to as the baseline. The second experiment uses the phonetic transcription obtained from the Wav2Vec 2.0 model, which we call the Wav2vec2-based model. The third experiment relies on the phonetic segmentations obtained from forced alignment using Astali, referred to as the Astali-based model. Finally, the fourth experiment uses segmentations manually corrected by an expert, which we will call the Expert-corrected model.

\subsection{model parametres}
All models were trained for 300 epochs with a batch size of 10. 
Training was performed using the Adam optimizer with a learning rate of $10^{-3}$, and early stopping was applied with a patience of 10 epochs on the validation set. Silences between sentences were removed, resulting in a total of 45,000 frames. The dataset was randomly shuffled and split into 80\% training, 10\% validation, and 10\% test subsets. The same randomization was used across all experiments, and all implementations were carried out in PyTorch.

\section{Results}
\begin{table*}
\caption{Comparison of RMSE (mm) and MEDIAN (mm) for different input features}
\label{tab:input_comparison}
\centering
\resizebox{1\linewidth}{!}{
\begin{tabular}{|l|c|c|c|c|c|c|c|c|}
\cline{2-9}
\multicolumn{1}{c|}{} & \multicolumn{2}{c|}{\textbf{baseline}} & \multicolumn{2}{c|}{\textbf{Wav2vec2-based model}} & \multicolumn{2}{c|}{\textbf{Astali-based model}} & \multicolumn{2}{c|}{\textbf{Expert-corrected model}}\\ 
\cline{2-9}
\multicolumn{1}{c|}{} & \textbf{RMSE} & \textbf{MEDIAN} & \textbf{RMSE} & \textbf{MEDIAN} & \textbf{RMSE} & \textbf{MEDIAN} & \textbf{RMSE} & \textbf{MEDIAN}\\
\hline
\textbf{Arytenoid}           & 1.63 $\pm$\, 1.02 & 1.38 & 1.83$^{*}$ $\pm$\ 1.13 & 1.55 &1.88$^{*}$ $\pm$\ 1.17 & 1.58 & 1.81$^{*}$ $\pm$\ 1.09 & 1.54\\
\textbf{Epiglottis}          & 1.52 $\pm$\, 0.88 & 1.33 & 1.71$^{*}$ $\pm$\ 0.99 & 1.49 &1.77$^{*}$ $\pm$\ 1.04 & 1.53 & 1.71$^{*}$ $\pm$\ 0.99 & 1.48\\
\textbf{Lower lip}           & 1.53 $\pm$\, 0.83 & 1.35 & 1.67$^{*}$ $\pm$\ 0.96 & 1.44 &1.64$^{*}$ $\pm$\ 0.99 & 1.40 & 1.54$^{*}$ $\pm$\ 0.88 & 1.34\\
\textbf{pharyngeal}          & 1.08 $\pm$\, 0.58 & 0.96 & 1.22$^{*}$ $\pm$\ 0.67 & 1.08 &1.23$^{*}$ $\pm$\ 0.70 & 1.07 & 1.19$^{*}$ $\pm$\ 0.65 & 1.04\\
\textbf{Velum}               & 1.34 $\pm$\, 0.67 & 1.21 & 1.42$^{*}$ $\pm$\ 0.70 & 1.29 &1.37$^{*}$ $\pm$\ 0.68 & 1.23 & 1.34$^{*}$ $\pm$\ 0.66 & 1.21\\
\textbf{Tongue}              & 2.33 $\pm$\, 1.17 & 2.10 & 2.53$^{*}$ $\pm$\ 1.36 & 2.24 &2.57$^{*}$ $\pm$\ 1.38 & 2.26 & 2.45$^{*}$ $\pm$\ 1.25 & 2.18\\
\textbf{Upper lip}           & 1.15 $\pm$\, 0.55 & 1.03 & 1.32$^{*}$ $\pm$\ 0.66 & 1.18 &1.19$^{*}$ $\pm$\ 0.59 & 1.07 & 1.16$^{*}$ $\pm$\ 0.57 & 1.04\\
\textbf{Vocal folds}         & 1.49 $\pm$\, 0.84 & 1.31 & 1.68$^{*}$ $\pm$\ 0.96 & 1.48 &1.77$^{*}$ $\pm$\ 1.05 & 1.53 & 1.68$^{*}$ $\pm$\ 0.97 & 1.46\\ \hline
\textbf{Mean}                & 1.51 $\pm$\, 0.91 & 1.30 & 1.67$^{*}$ $\pm$\ 1.03 & 1.44 &1.68$^{*}$ $\pm$\ 1.07 & 1.42 & 1.61$^{*}$ $\pm$\ 0.99 & 1.37\\
\hline
\end{tabular}
}
\\[1pt]$^{*}$Significant difference compared to the baseline result ($p < 0.05$) based on a t-test.
\end{table*}

Table \ref{tab:input_comparison} compares the performance of the four experiments based on the type of input features used: baseline, automatic phonetic transcription (Wav2Vec 2.0), phonetic segmentation obtained through forced alignment (Astali), and manually corrected segmentation. All models were evaluated using the RMSE and median, both expressed in millimeters. Additionally, a Student's t-test was performed to assess the statistical significance of the observed differences.

All models converged within 61 epochs. The baseline model using MFCCs achieved the lowest mean RMSE (1.51 mm), outperforming the Wav2vec2-based (1.67 mm), Astali-based (1.68 mm), and Expert-corrected (1.61 mm) models. When examining results per articulator, the baseline model obtained the best performance for seven articulators, with the exception of the velum, where the Expert-corrected model slightly outperformed it.

A similar pattern is observed for the median values: the baseline model consistently achieved lower medians on average and for seven articulators. The Expert-corrected model ranked second, showing better performance than the Wav2vec2-based and Astali-based models, which performed similarly, although the Wav2vec2-based model was slightly better than the Astali-based model on certain articulators.

\begin{figure}[ht] 
  \centering
  \includegraphics[width=0.475\textwidth]{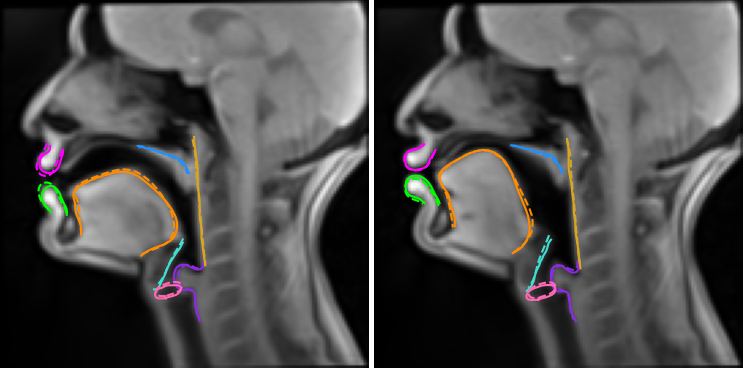}
 \caption{
    Example of two inversions compared to the original contours tracked in the rt-MRI film from baseline model.  
    The dotted lines show the predicted contours, while the solid lines represent the original ones. The RMSE is 1.36 mm for the phoneme ’a’ on the left and 1.50 mm for ’t’ on the right.   
}
  \label{fig:inference} 
\end{figure}

Figure \ref{fig:inference} illustrates an example of inference using the baseline model. The RMSE is 1.36 mm for the phoneme /a/ on the left and 1.50 mm for the phoneme /t/ on the right.


\section{Discussion}
The results show that the baseline model outperforms the other models, achieving a mean RMSE of 1.51 mm and a mean median error of 1.30 mm. This indicates that directly using the speech signal is more effective than relying on phonetic segmentation. These results can be explained by the fact that MFCCs primarily capture both the spectral and dynamic structure of the speech signal.

In contrast, phonetic segmentation relies on discrete linguistic units, which leads to a loss of essential acoustic information for the accurate prediction of articulatory contours. Moreover, this approach is sensitive to errors related to transcription or forced alignment. Even when manually corrected, phonetic segmentations remain limited by this discrete representation and by the inherent ambiguities of the phoneme–articulation relationship, which is underdetermined.

As a result, passing through a phonetic representation introduces an excessive simplification of the speech signal, thereby limiting model performance compared to a continuous acoustic approach based on MFCCs. When comparing approaches that rely solely on phonetic segmentation, the model using expert-corrected annotations outperforms the two other phonetic-based models, highlighting the critical role of segmentation accuracy in improving articulatory prediction performance. In addition, the Wav2vec2-based model, which relies on automatic transcription, slightly outperforms the Astali-based alignment model. This difference can be explained by the nature of the input representations: the Wav2vec2-based model uses phoneme posterior probability distributions, which preserve a degree of uncertainty and temporal smoothness, whereas the Astali-based model relies on a hard phoneme representation encoded as one-hot vectors, resulting in a more discrete and less informative input.

\section{Conclusion}
In this work, we compared several approaches by analyzing the impact of different levels of phonetic segmentation accuracy starting from a denoised speech signal. The results show that the baseline model based on continuous acoustic representations, in particular MFCCs, remains the most effective for vocal tract contour reconstruction, achieving the lowest RMSE and median error. This confirms the importance of fine spectro-temporal information in the speech signal for articulatory contour prediction.

Approaches relying on phonetic segmentation, although they provide a more direct linguistic interpretation and reduce dependence on raw acoustic features, are inherently limited by the discrete nature of phonemes and by the loss of intra-phonemic and coarticulatory information. As a consequence, their overall performance remains below that of the acoustic baseline.

Nevertheless, the improvements observed with manually corrected phonetic segmentations highlight the critical role of temporal and phonetic annotation accuracy. Among phonetic-based approaches, expert-corrected annotations after alignment lead to significantly better results than fully automatic segmentations. Furthermore, the results emphasize the advantage of probabilistic phoneme representations, which preserve part of the uncertainty and temporal continuity of the speech signal, compared to strictly discrete one-hot encodings.

\clearpage
\bibliographystyle{IEEEtran}
\bibliography{ref}
\end{document}